\documentclass[11pt]{article}
\usepackage{epsf}
\usepackage{mcite}
\newcommand{\fg}[2]{\null\centerline{\epsfxsize=#1truecm\epsffile{#2}}}

\newcommand{\beq}{\begin{equation}}
\newcommand{\bfg}[1]{\begin{figure}[#1]}
\newcommand{\efg}{\end{figure}}
\newcommand{\eeq}{\end{equation}}
\newcommand{\bary}{\begin{eqnarray}}
\newcommand{\rc}{\nonumber\\}
\newcommand{\eary}{\end{eqnarray}}

\newcommand{ \eq}{\ =\ }
\newcommand{\moy}[1]{\mbox{$\langle #1 \rangle $}}

\newcommand{\bigf}{I\!\! F}

\newcommand{\expp}[1]{\exp{\Big(#1\Big)}}
\newcommand{\as}{\alpha_s}

\newcommand{\go}{\gamma_0}

\newcommand{\psm}{\protect\small}
\begin{document}

\bibliographystyle{phreportmc}

\title{
 Energy Conservation Constraints on Multiplicity Correlations in
 QCD Jets 
\thanks{Nice INLN 96
/01; Saclay T96/002; work supported in part by the European Community Contract CHRX-CT93-0357 }\\}

\author{J.-L. Meunier\thanks
{I.N.L.N., Universit\'e
de Nice-Sophia
Antipolis, Unit\'e Mixte
de Recherche du CNRS, UMR 129, 1361 Rte. des Lucioles 06560 Valbonne, France}
\ and R. Peschanski\thanks{CEA, Service de Physique Th\' eorique, CE-Saclay
F-91191 Gif-sur-Yvette C\' edex, France}}
\date{}
\maketitle
\begin{abstract}
{
We compute analytically the effects of energy conservation on the 
self-similar structure of parton correlations in QCD jets. 
The calculations are performed both in the constant and running 
coupling cases. It is shown that the corrections are 
phenomenologically sizeable. On a theoretical ground, energy conservation constraints  preserve the scaling properties of correlations in QCD jets beyond the leading log approximation.
}
\end{abstract}
\vskip 1truecm

\section{Introduction}
The structure of parton multiplicity correlations within QCD jets 
and in particular their self-similar properties 
have been recently studied\cite{DD93,BMP,*ISMD92,OW93}. The main result is twofold~:
\begin{itemize}
\item The structure of a QCD jet  is {\it multifractal}, in the sense that
the multiplicity density of gluons does not occupy uniformally the available 
phase-space when it is computed   in the Double-Leading-Log 
approximation (DLA) of the perturbative QCD expansion. More precisely,
if one measures the inhomogeneities of the multiplicity distribution using the scaled factorial moments of order $q$, one writes~:
\begin{equation}
\label{inter}
{\cal F}_q(\Delta)\equiv {\moy{n(n-1)..(n-q+1)}_\Delta \over \moy{n}_\Delta^q} \propto \Delta^{(q-1)(1-{\cal D}_q/d)}
\end{equation}
where $n$ is the multiplicity of partons registered in a small phase-space interval
$\Delta.$ ${\cal D}_q $ is called the {\it fractal} dimension of rank $q$
of the density of gluons while $d$ is the overall dimension of the phase space under 
consideration (in practice, $d=2$ for the solid 
angle, and $d=1$ if one has integrated over, say the
azimutal angle, by keeping fixed the opening angle with respect to the jet axis). The multiplicity distribution is uniform if $
{\cal D}_q \equiv d,$ while it is fractal if it is smaller. Then {\it
multifractality} is for a $q-$dependent dimension.

In the DLA, assuming a fixed QCD coupling constant $\as$, the result is the following:
\begin{equation}
\label{diminter}
{\cal D}_q\eq\go\ \frac{q+1}{q}
\end{equation}
where $\go$, considered to be small, is given by~:
\beq
\label{go}
\go^2 \eq 4 N_c\ \frac{\as}{2\pi} \ ;\ \ \ N_c=3
\eeq
\item In the running coupling constant scheme, the jet structure is modified
by scaling violation effects. It leads to a multifractal dimension slowly varying with 
the angular variables of observation. Furthermore, one observes\cite{BMP}, at small angles ${\cal D}_q \to d,$ i.e. a dynamical saturation of multifractality.
The calculation gives~:
\beq
\label{inter5}
{\cal D}_q(z)
\simeq \go\left(\as(E\theta_0)\right) \frac{1+q }{q}\ \frac{2}{1+\sqrt{1-z}}
\eeq
where the  scaling variable $z$ is defined as: 
\beq
\label{Scal}
z=\frac{\log{\theta_0/\theta}}{\log{E\theta_0/\Lambda}}.
\eeq
Note that the fractal dimension is no more constant and depends
on both the angular direction($\theta_0$) and angular aperture ($\theta$) of the observation window. In practice, the {\it fractal} dimensions have a tendency to increase with $z$ and $\go$ and to reach the {\it saturation} point where
they become equal to the full dimension $d.$ This saturation effect comes
from the increase of the coupling constant at larger distances and
signals the onset of the non-perturbative regime of 
hadronization.
\end{itemize}
We want to reconsider these results by going beyond the approximations
made in Refs.\cite{DD93,BMP,OW93}. Some results exist which include  
Next-Leading-Log corrections to the anomalous dimensions\cite{DD93}, but
they do not include the energy-conservation constraints. Indeed, it is well known that pertubative QCD resummation in the leading log approximation (DLA) predict  too strong global multiplicity moments, at least when  compared to the hadronic final state observed in experimental data on jets\cite{Tran}. In fact, as pointed out in Ref.\cite{D93}, the energy conservation (EC) constraint at the  triple parton vertices may  
explain  the  damping of multiparticle moments observed in the data. While the EC correction is perturbatively of higher order ($\as^{3/2}$ as compared to $\sqrt{\as})$ and gives a rather moderate effect on the mean multiplicity, the calculation of  multiplicity moments of higher rank $q$ gives a contribution of order $q^2\as^{3/2}$ which become important already  for $q=2.$ As we shall demonstrate, this effect is even more important for  local correlations in phase-space.

The main goal of our paper is the analytical calculation of
the EC effects on the fractal dimensions ${\cal D}_q$
in the one and two-dimensional angular phase-space. We show that the self-similar structure of correlations is preserved beyond DLA and we give their analytic expression.
As an application we calculate the EC effects on the fractal dimensions of QCD jets  produced in $e^+\!-e^-$ collisions at  LEP energy\cite{OPAL92,ALEPH92,DELPHI92}.
We find that these effects cannot be neglected either in the phenomenological analysis
or in the theoretical considerations based on the perturbative QCD expansion.
Under the EC effect, the fractal dimensions increase and, in the case of a running $\alpha_s,$ strengthen the saturation phenomenon. 
In particular, the fractal dimension ${\cal D}_2$ becomes now of  order unity, in agreement with both numerical QCD simulations and experimental data. Note, from the theoretical point of view, the interesting interplay which appears with the property of KNO scaling of multiplicity 
distributions\cite{KNO}.
 
The plan of the paper is the following. In  the nextcoming section 2, we recall the QCD formalism for the generating function of multiplicity moments
for the {\it global} jet multiplicity, formulate the EC problem and give an explicit solution for the first global moments.
Section 3 is devoted to the {\it local} multiplicity study of QCD jets, i.e. the correlations/fluctuations in small angular windows (angular intermittency\cite{BMP}). We compute
the fractal dimensions and compare our analytical results (including rather significant EC effects) both with the relevant experimental data and with a computer simulation of QCD jets\cite{DJLM,*Cracow}. Our conclusions can be found
in section 4.

\section{The Global Jet multiplicity distribution}
We will use the following notations:
${\cal Z}(u,Q)$ stands for the QCD generating function of the global factorial multiplicity moments $\bigf_q$ (for a gluon of virtuality Q decaying into gluons).     The total jet 
multiplicity (first global moment) is denoted $N(Q).$
\beq
\label{notations}
\bigf_q\ \equiv {\moy{n(n-1)..(n-q+1)}}_{jet} \eq\ \frac{\partial^q \cal Z}{\partial u^q}\big|_{u=1}.
\eeq
When one takes into account energy conservation at the fragmentation vertex, 
${\cal Z}$ is governed by an evolution equation which can be sketched as a classical fragmentation mechanism:
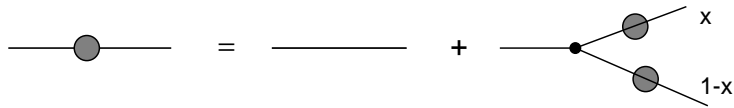
\bfg{thb}
\fg{10}{master1.ai}
\efg

\noindent where the black points represent the 
generating function of the jet and its sub-jets 
while the arrows stand for  the jet and sub-jet axis directions.
One obtains the well-known QCD evolution equation\cite{Tran} ~:
\bary 
\label{Z}
&\displaystyle{\frac{\partial {\cal Z}(u,Q)}{\partial \log(Q}} &\eq \int_0^1 \go^2\frac{dx}{x}\Big|_+{\cal Z}(u,Qx)\ {\cal Z}(u,Q(1-x))\rc
&&\eq \int_0^1 \go^2\frac{dx}{x}\left[{\cal Z}(u,Qx){\cal Z}(u,Q(1-x))-{\cal Z}(u,Q) \right]
\eary
where $\frac{dx}{x}\big|_+$ is the principal-value distribution coming from the triple gluon vertex and $\theta$ is the angular aperture of the jet which plays the r\^ole of a time variable. The energy of the primordial parton is $E$ and its virtuality is $Q\simeq E\theta$.

Note that this equation can be enlarged to include other non-leading-log
QCD contributions (the so-called Modified Leading-Log Approximation 
MLLA\cite{Tran}).
In fact an earlier study\cite{D93} has  shown that the EC effects
play the crucial role in correcting the global moments. We shall thus 
focus our discussion on the EC effects
but our method can be enlarged without major problems to the full MLLA equation.
\subsection{The mean multiplicity}
The solution for the mean multiplicity, $N(Q)$ is known\cite{Tran,D93}.
let us rederive it in a language appropriate for further generalization to the correlation problem. 
Using Eq. (\ref{Z}), one obtains by differentiation:
\beq
\label{N}
\frac{\partial N(Q)}{\partial \log(Q)} \eq \int_0^1 \go^2 \frac{dx}{x}\left(N(Qx)+N(Q(1-x))- N(Q)\right).
\eeq
Let us now distinguish the frozen-coupling regime and that with running $\as.$

\begin{itemize}

\item{\it Frozen coupling constant}

In this case, Eq. (\ref{N}) can be solved using a power-like behaviour for $N$:
\beq
\label{Power}
N(Q)\eq N_0 Q^\gamma
\eeq
with the following link between $\gamma$ and the DLA value $\go$ 
of Eq.(\ref{go})~:
\beq
\label{gamma}
\go^2=\frac{\gamma^2}{1-\chi(\gamma)}
\eeq
The function $\chi$ is defined by:
$$
\chi(x)\ \eq x \left({\Psi(1+x)-\Psi(1)}\right)
$$
where  ${\displaystyle \Psi(z)={\partial \log{\Gamma(z)}\over \partial z}}.$
When $x \to 0,$  $\chi \simeq x^2$, $\gamma \simeq \go-\go^3/2$ , which confirms that
the EC correction to  $N$ is a (${\cal O}(\as^{3/2})$) correction.
However, this correction appears in the exponent of the multiplicity and is thus  not negligeable at LEP energies. In practice, it "renormalizes" the multiplicity
exponent, see Fig.\ref{gog}.
At LEP energy ($E = M_{{\cal Z}_0}$), the correction can reach 20\%.
Note that $\chi$ 
can be numerically replaced by $x^2$ up to $x=1.5$.
\bfg{thb}
\fg{10}{gog.ai}
\caption{
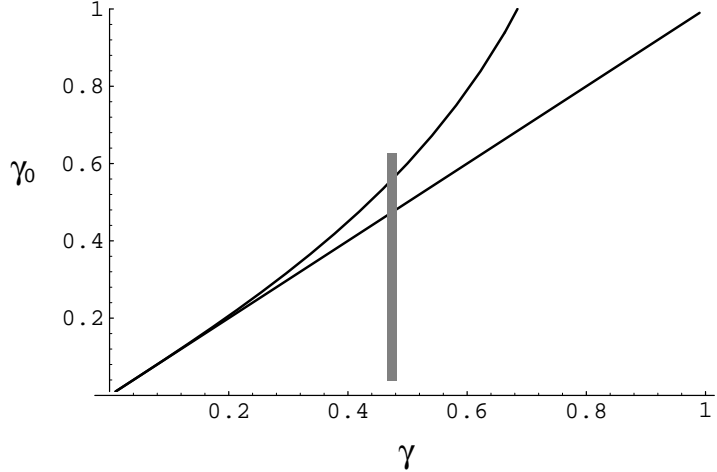
\psm{\bf a.}$\go$ as a function of the "renormalized"
 value $\gamma$; The grey band corresponds to the average value of 
$\go$ at LEP.}
\label{gog}
\efg
\item {\it Running coupling constant}

Let us now reconsider equation (\ref{N}) when $\alpha_s$ is running.
One has:
$$
\go^2 \eq 4 N_c{\as(Q)\over 2 \pi}\eq {c_0^2\over\log{(Q/\Lambda)} }\ ;\ \ c_0\eq\frac{6}{\sqrt{33-2n_f}}, 
$$
where $n_f$ is the number of flavors ($n_f$ = 5 at LEP).
The solution of Eq.(\ref{N}) is of the form  $N(Q) = N_0 \expp{2c\sqrt{\log (Q)}}$, where $c$ is a constant. Using the following general identity,
\bary
\label{Gi}
&N(Q f(x))/N(Q)\simeq 
\expp{2c \sqrt {\log{Qf(x)/\Lambda}}
-2c \sqrt {\log{Q/\Lambda}}}\rc
&\simeq  
\expp{c\ \log f(x)/\sqrt {\log{Q/\Lambda}}}\eq \left[f(x)\right]^{\ c /\sqrt{\log{Q/\Lambda}}}.
\eary
one obtains from Eq.(\ref{N}):

\bary
\label{RN}
&N(Q)\simeq N_0\expp{2 c\sqrt{\log{Q/\Lambda}}}\rc
&c^2\simeq c_0^2\left(1-\chi{(\gamma  )}\right),
\eary

where $\gamma \equiv c/\sqrt{\log{Q/\Lambda}}.$   As already pointed out, $\chi(\gamma) \simeq \gamma^2$ when
$\gamma$ is small, that is  the EC effect on the exponent of the
mean multiplicity is a non-leading correction. However, as in the frozen coupling case, it "renormalizes" the
behaviour of the multiplicity.
\end{itemize}
\subsection{The global second moment}
If one goes to the second derivative of  equation (\ref{Z}), one obtains the evolution equation for the global factorial moment, $\bigf_2$~:

\beq
\label{GF2}
\frac{\partial \bigf_2 (Q)}{\partial \log(Q)} \eq \int_0^1 \go^2\frac{dx}{x}\left(\bigf_2(Qx)+2 N(Qx) N(Q(1-x))+\bigf_2(Q(1-x))-\bigf_2(Q)\right).
\eeq

The way of dealing with this equation is to assume 
that $ \bigf_2(Q)/N^2(Q)$ is a slowly varying function of
$Q.$ This property, which is one of the KNO scaling
relations $ \bigf_q(Q)\propto N^q(Q),$ is known to be correct for 
the moments predicted by QCD\cite{Tran}.  Using this structure together with the power-like behaviour of $N $, (see Eq. (\ref{Power})) gives~:
\beq
\label{F2}
\bigf_2=\left(\frac{2\Gamma^2(\gamma)}{\gamma\Gamma(2\gamma)(3+\chi(2\gamma)-4\chi(\gamma))} \right)\left( N(Q)\right)^2.
\eeq
The result is displayed on Fig.\ref{f2b}. The EC effects are
 clearly seen as a serious damping of the KNO ratio $\bigf_2/N^2$. 
\bfg{thb}
\fg{10}{f2b.ai}
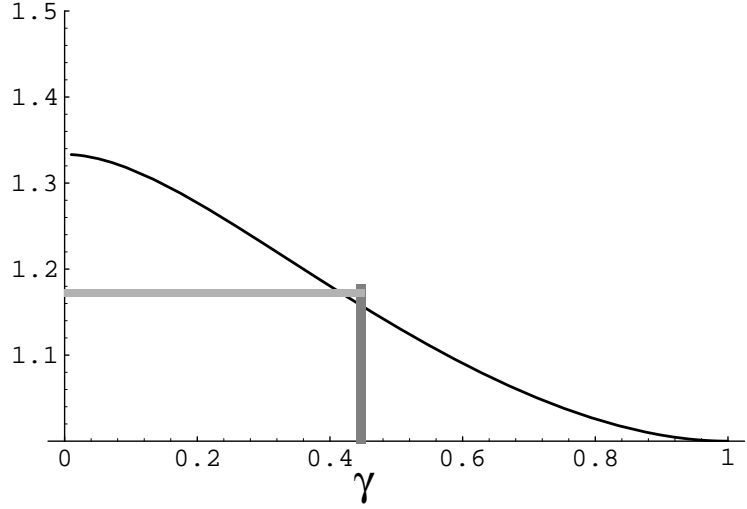
\caption{\psm $\bigf_2$ as a function of $\gamma$. The DLA result (4/3) is  obtained when $\gamma\to 0$. The vertical grey band is for $\go$ at LEP , while
the horizontal one corresponds to the experimental value quoted in ref. \protect\cite{OPAL92} for one jet (in the forward hemisphere).}
\label{f2b}
\efg
The running of the coupling constant does not modify the result which is very close to the experimental data\cite{OPAL92} ($|bigf_2$ in one hemisphere).
One notices that Eq. (\ref{F2}) leads to a weak violation of KNO scaling due to the dependence of $\gamma$ on $Q.$ 
\section{The Local Jet multiplicity distribution}
Let us come now to the main topics of our paper, namely the computation of the local correlations between partons.
The characteristic feature of the jet multiplicity structure is that the evolution 
equation for the local density of partons is linear. It 
can be deduced from the branching structure of jet fragmentation in QCD.
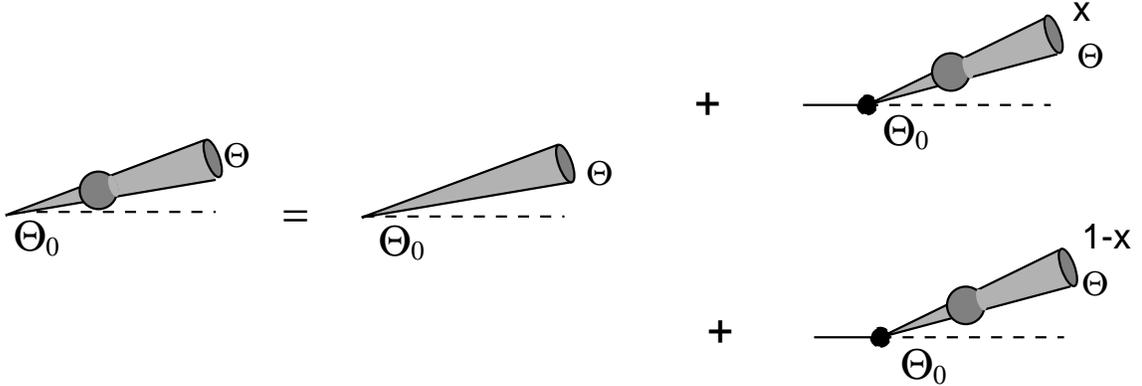
\bfg{thb}
\fg{16}{master2.ai}
\efg

Let us denote by $H(u,Q)$ the generating function for the  {\sl density} of factorial moments:

\beq
\label{density}
H_q\equiv\frac{\partial ^q H}{\partial u^q}\Big|_{u=1}\eq\frac{{F}_q(\Delta)}{\Delta}\equiv \frac{{\moy{n(n-1)..(n-q+1)}}_\Delta}{\Delta}.
\eeq
The QCD evolution equation of the multiplicity in an
observation window of size $\theta$ pointing in a definite direction $\theta_0$ from the jet axis is schematically described in the figure. there are 2 contributions, one coming from the parton with a fraction $x$ of the available energy, the other  coming from the branch of the fragmentation process with $1-x.$ Once substracted the variation of the phase-space from $\theta_0$ to $\theta$ by defining moments of the {\it density} as in Eq.(\ref{density}), the only evolution comes from the energy degradation along the branch. In the DLA approximation, only one branch contributes to the hadron density evolution,  since one parton keeps essentially undisturbed by the branching. In the EC case where we do not neglect the recoil effect, the two branches contribute. Note that one has to take into account the corresponding loss of virtuality $Qx$ and $Q(1-x)$ respectively. One obtains~:
\bary 
\label{H}
&\displaystyle{\frac{\partial H(u,Q)}{\partial t}}
&\eq \int_0^1\go^2\ \frac{dx}{x}\Big|_+\left( H(u,Qx)+ H(u,Q(1-x))\right)\rc
&&\eq \int_0^1\go^2\ \frac{dx}{x}\left(\left[ H(u,Qx)-1\right]
+\left[ H(u,Q(1-x))-H(u,Q)\right]\right)
\eary
where $\log{\theta_0/\theta}\equiv t$ plays the r\^ ole of the evolution parameter  and the phase-space splitting depends on the dimensionality, namely $\Delta\propto \theta^d$. One branch of the iteration contributes to
  $H(u,Qx)$ and the other to $H(u,Q(1-x))$ in the integrand, while the negative contributions
are necessary to ensure the finiteness and  unitarity conditions,
namely $H(1,Q) \equiv 1$ and $\sum_\Delta {\partial H(u,Q)}/{\partial u}\big|_{u=1} = \sum_\Delta
F_1(\Delta) \equiv N(Q).$

The solution of Eq.(\ref{H}) is obtained by using the KNO scaling property of the multiplicity distribution inside a cone of fixed angular aperture $\theta$.
In terms of the generating function $H$, it writes:
\beq
\label{KNO}
H(u,\lambda Q)\eq H(uN(Q\lambda)/N(Q),Q)
\eeq
Inserting the property (\ref{KNO}) into Eq.(\ref{H}), and performing the $q^{th}$ derivative, one gets:
$$
\frac{\partial H_q}
{\partial t} \eq H_q \ \int_0^1 \go^2 \frac{dx}{x}\left(\left(
{\frac {N(Qx)}{N(Q)}}\right)^q
+\ \left({\frac {N(Q(1-x))}{ N(Q)}}\right)^q - 1\right) 
$$
\bary
\label{Hq}
 \eq H_q \ \int_0^1 \go^2 \frac{dx}{x}\left(
x^{q\gamma} +\ 
(1-x)^{q\gamma} - 1\right),
\eary
where we have used Eq.(\ref{Gi}). Notice that Eq. (\ref{H}), as well as (\ref{Z}), can be understood in terms of classical fragmentation models\cite{BMP}. The connection between the present formalism and classical fragmentation is also a consequence of the KNO relation (\ref{KNO}). 

As a check of this procedure, let us recover the known results 
of the DLA approximation\cite{BMP}. Neglecting the EC terms in Eq.(\ref{H}), one obtains~:
\beq
\label{DLAH}
\frac{\partial H(u,Q)}{\partial t}
\eq \int_0^1\go^2\ \frac{dx}{x}\left( H(u,Qx)-1\right),
\eeq
from wich one obtains the fractal dimensions (\ref{diminter}),(\ref{inter5}).
\subsection{ Frozen coupling } 
Equation (\ref{Hq}) reads~:
\beq
\label{Froze}
\frac{1}{H_q}\frac{\partial H_q}{\partial t}
\eq  {\go^2}\ \frac{(1-\chi(q\gamma))}{q\gamma}\eq  \frac{\gamma}{q}\ \frac{1-\chi(q\gamma)}{1-\chi(\gamma)}
\eeq
where one makes use of relation(\ref{gamma}). 
 The corresponding corrections to the normalized factorial moments (\ref{inter}) and
their fractal dimensions can be  worked out~:
\beq
\label{Ecdim}
{\cal D}_q
\eq \frac{\gamma }{q}\left[1+q+\frac{\chi(\gamma)-\chi(q\gamma)}{1-q}\right]
,
\eeq
 
\bfg{thb}
\fg{8}{ec_dims.ai}
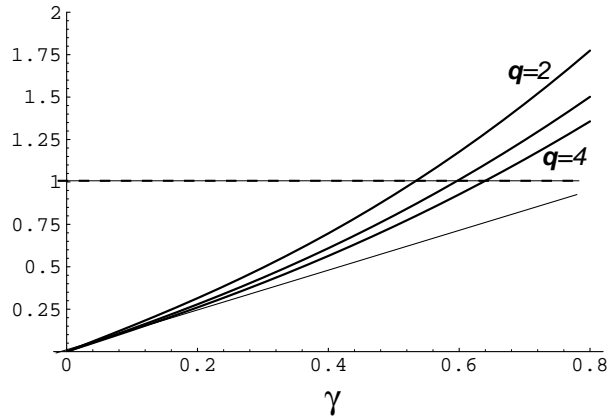
\caption{\psm The multifractal dimensions ${\cal D}_2, {\cal D}_3, {\cal D}_4 $ as functions of the coupling $\gamma$; The straight line is the DLA result for ${\cal D}_4$ as an example(\ref{diminter}). The saturation limit  is the dotted line at $d = 1$ .
}
\label{dim}
\efg

Notice that, while the correction to the unnormalized factorial moments,
see Eq. (\ref{Ecdim}), is 
a factor of
order $(1-q^2\gamma^2)/(1-\gamma^2)$ in the exponent and can be important,
 the
normalized ones are less sensitive to  EC corrections. Since  $\chi(\gamma)\simeq \gamma^2$ up to  $\gamma \simeq 1.5$, the fractal dimension
(\ref{Froze}) reads
\beq
\label{Apdim}
{\cal D}_q
\simeq  \frac{\gamma (1+\gamma^2)(q+1)}{q},\ \ q\gamma<1.5,
\eeq
leading to an enhancement  factor with respect to the multiplicity exponent at the same  order ($\as^{3/2}$).
However, when $q\gamma > 1.5,$
the modification becomes more important and $q-$dependent.
One observes an important increase of the fractal dimension, see 
Fig.\ref{dim}. 
\subsection{ Running coupling }
Let us consider again Eqs.(\ref{Hq}), (\ref{Froze}) with $\go$ defined with the running coupling and $\gamma$ as in formula (\ref{RN}). Neglecting both the variation of the coupling in the EC correction factors (higher order terms) and the $x$-dependence of the coupling (this approximation have been shown to be quite right in  the   range ($z<.6$) in ref. \cite{BMP}), one obtains~:
\beq
\label{Rundim}
{\cal D}_q(z)
\simeq  \frac{\gamma }{q}\left[1+q+\frac{\chi(\gamma)-\chi(q\gamma)}{1-q}\right]\ \frac{2}
{1+\sqrt{1-z}},
\eeq
where $\gamma$ is expressed as in formula (\ref{gamma}). Notice that the saturation effect is stronger due to the running of the coupling (see, e.g., Fig.\ref{ecrun} for the one-dimensional case). 
\bfg{thb}
\fg{13}{ecrun.ai}
\caption
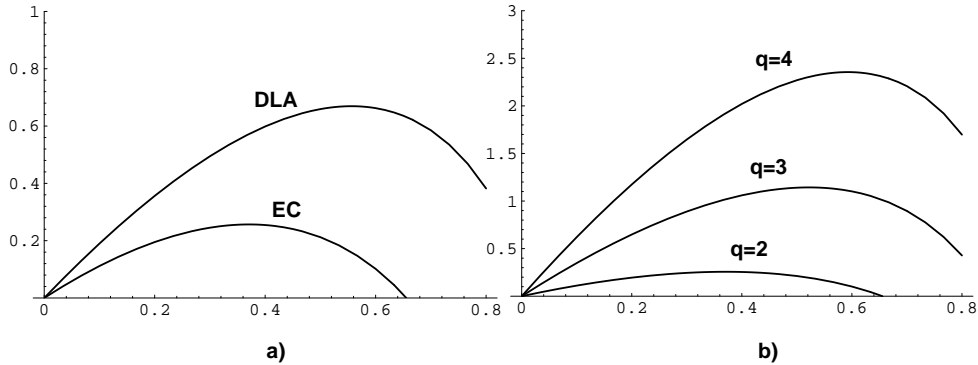
{ $\log[{\cal F}_q(z)/{\cal F}_q(0)]$, as a function of the scaling variable, $z=\log[\theta_0/\theta]/\log[E\theta_0/\Lambda]$. {\bf a:} $q=3$, with and without EC corrections; {\bf b:} $q=2$ (lower), $q=3$ and $q=4$ (upper), including EC corrections; the value $\gamma=.44$ is consistent with the LEP value of $\as$.
}
\label{ecrun}
\efg

As a phenomenological application of our results, we perform a comparison of our predictions with  experimental 
results, which while yet preliminary, have been presented as a first detailed analysis of correlations in a jet. As stressed in the experimental paper\cite{Vietri}, the comparison of analytical QCD predictions with data on correlations is not easy due to the  hadronization effects. The best candidate for a comparison is the study of particle correlations in a ring 
$\theta_0 \pm \theta$ around the jet axis. The variation on $\theta$ allows one to obtain the dependence in $z,$ while considering as an observable the ratio  $\log[{\cal F}_q(z)/{\cal F}_q(0)]$ minimizes the hadronization corrections\cite{BMP}.
 
The results are displayed on   Fig. \ref{McDat}. The 
result of a Monte-Carlo simulation of the QCD parton cascade \cite{DJLM,*Cracow} is also shown on the same plot. The simulation is useful to take into account
the corrections with respect to analytical calculations which occur when the parton cascade is correctly reproduced, e.g. overlap effects between angular domains, subasymptotic energy-momentum effects, hadronization effects (in part) etc.... For  this comparison we make use of  the following parameters : $\as(E\theta_0) =.135 $ for $\theta_0=25^\circ $(corresponding to $\as=.12$ for the whole jet with $Q_0=45$ Gev). This leads to  $\gamma=.44$ or, from Eq.(10),
 $\go =.5$.
\bfg{thb}
\fg{10}{ec_mc_exp.ai}
\caption
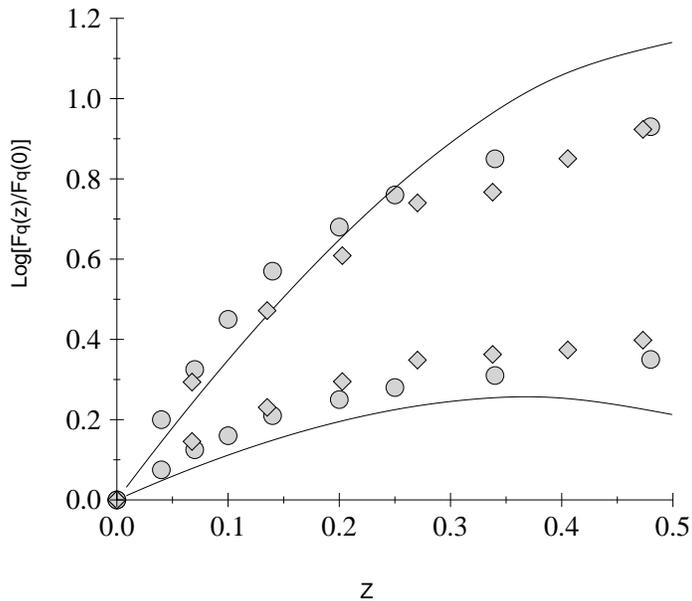
{\psm  Phenomenological predictions for $\log[{\cal F}_q(z)/{\cal F}_q(0)]$ as a function of the scaling variable $z$.
Upper curves,triangles and circles: $q=3$, lower ones $q=2$.
Continuous lines: the analytical calculations from \protect\ref{inter5}; circles: Preliminary Delphi data from Ref.\protect\cite{Vietri}; squares: Numerical data from the Monte-Carlo simulation\protect\cite{DJLM,*Cracow}}
\label{McDat}
\efg

The analytical QCD predictions happen to be  in reasonable agreement with both experimental results  and numerical Monte-Carlo simulations, taken into account the approximations considered in the calculation. In particular, some bending of the factorial moments at small aperture angle $\theta$ is qualitatively reproduced. This effect is a typical predictions of the running coupling scheme, which is thus favoured by the data with respect to the frozen coupling case.

 The full study of hadronization effects remain to be done. However, we may 
 notice that when the slope of the intermittency indices (Fig. \ref{ecrun}) goes to zero, the effective coupling $\gamma$  is around $1$ and one  enters   the not well known strong coupling regime  of QCD. As a matter of fact, the only free parameter  left in the Monte-Carlo simulation is the effective cut-off of the parton cascade i.e. the maximum value $\as,\ \alpha_0$, one considers  in the development of the parton branching process. In the example considered in the figure, we have found $\alpha_0\simeq .54,$ which is a quite reasonable value for  switching on hadronization. All in all it is a positive fact  that Local Parton-Hadron Duality seem to work reasonably for such refined quantities as correlations, once some caution is taken to minimize non-perturbative QCD corrections.

\section{Conclusions}
The Energy Momentum Conservation effects (or recoil effects) on the correlation properties of QCD jets are thus under some analytical control. The method proposed to handle these effects is the search for scaling solutions of the QCD generating function of {\it local} multiplicity density moments in angular phase-space. It is interesting to note the deep connection between the KNO-scaling\cite{KNO}  and the self-similar properties of correlations in the jet. This structure is reminiscent of general properties of fragmentation models and could be more fundamental than the level of approximation used to derive the result from  QCD.  Indeed, the same qualitative result is expected from the solution of the equation including EC terms plus other dominant next-to-leading effects (e.g. MLLA approximation scheme, see ref.(\cite{Tran})).

The phenomenological outcome  of this study is that the analytical predictions come close to the experimental data and confirmed by the numerical simulation. Parton-Hadron duality for correlations seems to be  supported  provided hadronization effects can be minimized. Thus the self-similar structure of the QCD branching processes may eventually emerge from the background in the experimental analysis. Hadronization becomes dominant when the factorial moments bend and acquire a zero slope ($z\simeq 0.5$ in terms of the scaling variable of formula (\ref{Scal})). If one would insist to increase $z$ by reducing  the opening angle of the detection window,
 one would find  properties  closer to soft processes, and thus difficult to predict theoretically. 

From the theoretical point of view, the partial cancellation  of EC corrections observed in the scaled factorial moments, leading to a moderate correction  to the behaviour determined by the leading-log approximation, is a consistency check   of the perturbative resummation predictions. Note however that this remark does not hold for moments of high rank $q\ge 4$, where a stronger correction is expected. The question remains to know whether this indicates a limitation of the perturbative calculations or the signal of a dynamical  mechanism.

\end{document}